\documentclass{eas33}
\usepackage{graphicx}
\TitreGlobal{2nd ARENA Conference: The Astrophysical Science Cases at Dome C}
\begin{document}

\title{Motivations for Imaging Spectroscopy at~Dome~C}
\author{A. Kelz}
\address{Astrophysikalisches Institut Potsdam,
		An der Sternwarte 16,
		14482 Potsdam, Germany
		\email{akelz@aip.de}}
%
%
\begin{abstract}
Antarctica offers unique conditions for ground-based observations, 
such as low sky background in the infrared, improved seeing, and low turbulence and scintillation noise. 
These properties are particularly beneficial for imaging, precision photometry and infrared observations. 
It may be less clear if Antarctica offers equally compelling advantages for spectroscopy, in particular in the optical domain. 
However, scientific programmes that make use of imaging (or 3D) spectroscopy for 
selected follow-up studies of IR surveys, long-term monitoring of 
extended targets and resolved stellar population studies in crowded fields, also benefit from the site conditions at Dome C.
\end{abstract}
\maketitle
%

\section{Scenarios for Imaging Spectroscopy}
The major motivations to conduct astronomical observations in Antarctica seem to be 
(wide-field) infrared surveys, diffraction-limited imaging and long-term accurate photometric monitoring. While for these programmes the conditions in space are superior, 
a ground-based site in Antarctica offers lower costs and the option to deploy larger 
telescopes and instruments. 
As spectroscopy requires larger collecting areas than imaging, it needs to be evaluated 
for which cases a median-sized telescope at Antarctica can outperform a larger one at temperate sites. 

However, if spectroscopy were combined with imaging, as in 3D- or Integral-Field Spectroscopy (IFS), which provides multiple spectra for each point of a 2-dimensional field, the case for Antarctica becomes more compelling. 
 
For extended targets, such as clusters, HII regions, 
nebulae, jets, or galactic environments, the use of integral-field techniques becomes 
necessary to overcome the source confusion in these complex fields. 
Integral field spectrophotometry is a useful technique for the study of resolved stellar populations in nearby galaxies (\cite{Roth2004}). The spectroscopy of individual extragalactic stars is limited by the accuracy to subtract the bright, spatially and wavelength-dependent non-uniform background of the underlying galaxy. 
Imaging spectroscopy is able to improve the limited accuracy of background subtraction that one would normally obtain with slit spectroscopy. 
Resolved stellar populations are foreseen targets for the 2nd generation 
instrumentation at the VLT such as MUSE (\cite{Bacon2006}), and provide a major science case for the planned ELTs (\cite{Hook2007}).  


A further case was made by Andersen (2007) for high-resolution imaging in the optical. 
Assuming a seeing of less than a 1/4 arcsecond for a significant fraction of the time, 
a 2.4m PILOT-like telescope (\cite{Storey2007}) with tip-tilt correction is diffraction limited. Such a facility could pickup some of the science cases of HST after its de-commissioning.  
This statement equally holds for imaging spectroscopy, as long as the spatial sampling 
of the integral-field-unit (IFU) does not degrade the input conditions. 
Most likely such an IFU will be based on an image slicer type, as this 
technique preserves the spatial information best. 
Image slicers at cryogenic temperatures are being used in SPIFFI; 
advanced image slicers are planed for MUSE, 
which will operate in the wavelength regime of 465-930nm. 
After the de-commissioning of HST, 
an integral-field spectrograph even at a median-sized telescope in Antarctica 
would be a unique facility for diffraction-limited spatial spectroscopy. 
Given its southern location, high resolution spectroscopic studies in 
the SMC/LMC are possible objectives.     


Many targets discovered in future IR-surveys will need spectroscopic follow-up  
and only a small fraction can be done from space. 
Given the wide-fields of IR-imaging surveys, integral field  spectroscopy over a large area would be desirable.  
A proposal was made by Maillard (2007) for a near-IR Imaging Fourier-Transform Spectrometer, that combines a wide-field with a high spectral resolution and can be used for Spitzer survey follow-ups.  
Less complicated, conventional IFUs can be used to study the details of complex  
star formation regions, for example in the Carina nebula or towards the Galactic Center (\cite{Zinnecker2007}). 


At Dome C, time-series data can be taken over months with duty cycles that are otherwise only possible using a network of telescopes. While the nights at the South Pole are longer, 
the seeing and the related scintillation noise are likely to be better at Dome C. 
Therefore, it shall be possible to obtain spectrophotometric data with greater precision than from any other ground-based site. 


The spatial position of gamma ray bursts or X-ray sources is only known with a limited accuracy of a few arcseconds. Here, imaging spectroscopy provides an increased error budget to ensure that the target is not missed altogether. 
The measured 2-dimensional PSF allows one to disentangle the spectra of the central source and the surrounding host galaxy. 


To detect and to measure the dynamics of low surface brightness objects, 
sufficient flux collection, rather than spatial resolution, is an issue. 
IFUs with high instrumental grasp, i.e. light collecting power (e.g. PPak-IFU, \cite{Kelz2006}) and spatial binning of spectra can be applied to improve the signal-to-noise ratio. 


For distant emission line sources (e.g. Ly-$\alpha$ galaxies) with a priori unknown redshifts, 3DS is the only technique that can reliable detect these. The 3D-data cube corresponds to a volume in space, which otherwise can only be recorded with time-consuming scanning techniques.

Given the above scenarios, two strawman models for imaging spectrographs may be considered:  an IFU with high (diffraction-limited) spatial resolution but a small field of view and one  
with high light collecting power over a larger field. 

\begin{table}[h!]
\begin{tabular}{l|l|l}
\hline 
characteristics 	& high resolution IFU 	& high-grasp IFU \\
\hline  
field of view		& several arcseconds	& few arcminutes  \\ 
spatial sampling	& $\approx 0.2$ arcsec	& coarse ($>1''$)\\ 
			& diffraction limited 	& high flux collection \\ 
IFU type		& image slicer  	& fiber bundle \\ 
			& 			& (+ OH-surpression) \\ 
spectrograph		& fixed setting 	& fiber-coupled \\ 
			& winterised at telescope & remote, bench-mounted \\ 
			& 			& multi-unit \\  
\hline 
application 		& crowded field spectroscopy & detection \& dynamics of \\ 
			& spectrophotometry & low surface brightness objects \\
			& resolved populations & star formation regions \\ 
\hline
\end{tabular}
\end{table}

\section{Synergies beyond Astronomy}
As a variety of scientific activities are being undertaken at Dome-C, 
it is beneficial to identify interdisciplinary links between those areas 
and astronomy. 
Perhaps the most obvious one is the overlapping interest between astronomical site 
testing and atmospheric research. The results from balloon flights to measure the 
temperature, pressure and wind profiles, and seeing measurements with telescopes 
are of common interest to both scientific branches. 
Apart from site testing, joint projects are in planing, such as the combination between 
ICE-T (AIP, \cite{Strassmeier2007}) and TAVERN (AWI) that uses a common infrastructure to measure the aerosol content of the atmosphere using a single star-photometer and a twin-telescope to photometrically monitor a stellar field. 
With respect to the above presented cases for astronomical spectroscopy, 
it is mainly the method and instrumentation that is well suited for other 
applications too. 
In particular, a fiber-coupled multi-channel spectrograph can not only be used for 
integral-field but also for multi-object spectroscopy. 
Instead of connecting the fiber input ends to a telescope, they may be equipped 
with optical sensors to measure certain physical parameters. 
Fibres can be deployed at locations that are otherwise difficult or not accessable 
to other sensor systems. Similar to ice-drilling programmes, sensor equipped 
optical fibres may be placed deep into the ice to measure pressures, temperatures, 
oxygen contents, etc.  
Hundreds of such fiber-sensors could be connected to a single multi-channel spectrograph, 
such as used for astronomical 3D-spectroscopy, and could yield important parameters, 
that are of interest for geophysics or glaciology.   
Fiber sensors that are distributed along a tower, to record atmospherical parameters, 
may help to determine the behaviour and size of the boundary layer in details and 
with improved spatial and temporal resolution.  

Apart from its main astrophysical interests, the AIP, together with the University of Potsdam, 
is actively engaged in research and development projects to advance methods in fiberoptical 
spectroscopy and sensing (see center for innovation competence: innoFSPEC-potsdam.de). 

\section{Conclusions}
In summary, scientific motivations for imaging spectroscopy in Antarctica are found 
i) in regions where spectral windows are present that are 
not accessible anywhere else from the ground (\cite{Burton2005});  
ii) for either time-consuming or frequent spectroscopic measurements, 
for which sufficient time from space will not be available; 
iii) for high-precission spectrophotometry that needs uninterrupted time series 
or 2D-monitoring (e.g. planets); 
iv) for imaging spectroscopy in the optical with high spatial resolution 
(e.g. resolved population studies, AGN/SN/GRB host galaxies). 
 
For the above applications, innovative imaging spectroscopy offers technical advantages, as compared to classical spectroscopy (\cite{Kelz2004}). 
It relaxes operational requirements (e.g. acquisition accuracy, dispersion compensation), which simplifies the instrumental design and reduces the number of movable parts as potential sources of failure. Fiber-coupled 
instruments can be placed remotely from the telescope in warm conditions.  
Finally, fiber-spectrographs, similar to the ones used in astronomical spectroscopy, 
may be coupled to advanced fiberoptical sensors and can provide data for a broad range of other scientific research areas, such as geophysics, environmental monitoring, 
atmospheric and climate research. Therefore, techniques and infrastructure  
that are being developed in Astrophysics 
could be used for other, interdisciplinary projects at Dome-C as well. \\ 

\noindent
AK gratefully acknowledges support from ARENA to attend this conference.



\begin{thebibliography}{99}
\bibitem[Andersen 2007]{Andersen2007} Andersen, M.I., 
2007, 2nd ARENA Conference, Potsdam (this issue) 

\bibitem[Bacon et al. 2006]{Bacon2006} Bacon, R., et al.\ 2006, 
Proc. SPIE, 6269 

\bibitem[Burton et al. 2005]{Burton2005}
Burton, M.~G. et al.\
2005, PASA, 22, 199

\bibitem[Hook 2007]{Hook2007} Hook, I.~M.\ 2007, EAS 
Public. Series, 25, 111 

\bibitem[Kelz et al. 2006]{Kelz2006} Kelz, A., Verheijen, M., Roth, M.~M. et al.
2006, PASP, 118, 129

\bibitem[Kelz 2004]{Kelz2004}
Kelz, A.
2004, AN, Vol. 325, Issue 6, p. 673


\bibitem[Maillard 2007]{Maillard2007} Maillard J.-P.,  
2007, EAS Publications Series, Volume 25, 2007, pp.245 


\bibitem[Roth et al. 2004]{Roth2004} Roth, M.~M., Becker, T., Kelz, A., Schmoll, J., 2004, ApJ, 603, 531


\bibitem[Storey et al. 2005]{Storey2007} Storey, J., Ashley, M., Burton, M., \& Lawrence, J.\
2007, EAS Public. Ser., Vol. 25, 255 

\bibitem[Strassmeier 2007]{Strassmeier2007} Strassmeier, K.G.,  
2007, EAS Publications Series, Volume 25, 2007, pp.233 


\bibitem[Zinnecker et al. 2007]{Zinnecker2007} Zinnecker, H., 
Andersen, M.~I., \& Correia, S.\ 2007, EAS Publications Series, 25, 183 



\end{thebibliography}
\end{document}